\documentclass[runningheads]{llncs}
\usepackage[T1]{fontenc}
\usepackage{amsmath, amssymb}
\usepackage{siunitx}
\usepackage{setspace}
\usepackage{graphicx}
\begin{document}
\title{Finite element models for Self-Deployable Miura-folded origami}
%
%\titlerunning{Abbreviated paper title}
% If the paper title is too long for the running head, you can set
% an abbreviated paper title here
%
\author{Suraj Singh Gehlot\inst{1}\and
Siddhanth Gautam\inst{1} \and
Sanhita Das\inst{1}}
\authorrunning{Sanhita Das et al.}
% First names are abbreviated in the running head.
% If there are more than two authors, 'et al.' is used.
%
\institute{Department of Civil and Infrastructure Engineering, Indian Institute of Technology Jodhpur, Jodhpur, Rajasthan - 342030\\
\email{sanhitadas@iitj.ac.in}\\
} 
\maketitle              % typeset the header of the contribution
\begin{abstract}
Origami-inspired self-deployable structures offer lightweight, compact, and autonomous deployment capabilities, making them highly attractive for aerospace and defence applications, such as solar panels, antennas, and reflector systems. This paper presents finite element frameworks for simulating Miura-origami units in ABAQUS, focusing on two deployment mechanisms: elastic strain energy release and thermally activated shape-memory polymers (SMPs). Validation against experimental data for elastic deployment demonstrates that the model accurately captures fold trajectories and overall kinematics. Parametric studies reveal the influence of hinge stiffness and damping on deployment efficiency. SMP-based simulations qualitatively reproduce stress-strain-temperature behavior and realistic shape recovery ratios. The study establishes that predictive numerical models can effectively guide the design of origami-based deployable structures for aerospace and defence applications, while highlighting the challenges associated with hinge modeling, damping effects, and thermomechanical actuation.

\end{abstract}

\section{Introduction}
Deployable structures are a cornerstone technology for space exploration, where mass, volume, and reliability are critical. Traditional mechanical deployment relies on hinges, actuators, or inflatables, but these are limited by mechanical complexity and reliability concerns. Origami-inspired designs exploit geometric folding patterns to achieve compact stowage and large deployed surface areas, making them particularly attractive for solar panels, antennas, and space telescopes. Among these, the Miura fold is notable for its single-degree-of-freedom deployment and structural efficiency. With the feature of self-deployability, it provides strong competition to roll-out and other deployment mechanisms.

The modeling of origami deployment can follow three primary approaches: (i) kinematic models, which address flat-foldability and rigid-panel geometry but ignore stresses and material behavior; (ii) bar-hinge or plate models, which introduce simplified stiffness parameters at creases but cannot resolve local stress fields or large nonlinearities; and (iii) finite element (FE) models, which are computationally expensive but capable of resolving detailed stress distributions, geometric and material non-linearities, and active-material (shape-memory-based alloys and polymers) actuation. This paper adopts the FE approach using ABAQUS \cite{smith2020abaqus} with the user-subroutine VUMAT for the constitutive model for shape-memory polymer. The present framework is capable of modelling two complementary mechanisms of deployment: elastic strain energy release through tape-spring analogues, and thermally activated recovery through shape-memory polymers. Two separate ABAQUS models are built addressing the two deployment mechanisms. For the elastic recovery, a four-panel Miura unit is folded elastically and released, and the panel trajectories are compared with the experiments in \cite{fulton2022deployment}. Further parametric studies for geometric parameters reveal crucial aspects of the deployment mechanisms. For the Shape Memory Polymer (SMP)-based recovery, a Miura column is subjected to compression and release under a thermal cycle, and a qualitative analysis of the shape recovery is performed. Finally, the article is concluded.

\section{Elastic Energy-based Deployment of a Single Miura Unit}
\begin{figure}[h]
    \centering
    \includegraphics[width=1\linewidth]{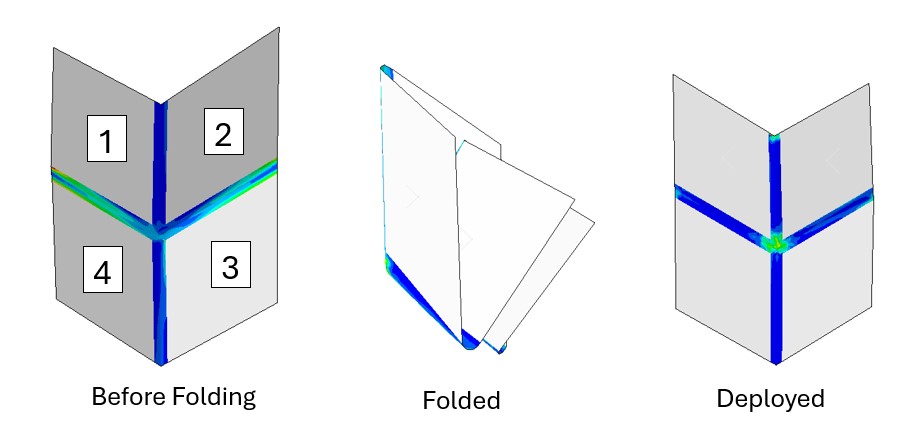}
    \caption{ABAQUS model of the 4-panel Miura unit in pre-folded, folded, and deployed states.}
    \label{fig:abaqus_model}
\end{figure}

A four-panel Miura fold unit with \SI{292}{mm} edge length panels was developed in ABAQUS in line with \cite{fulton2022deployment}. Each panel was modeled as a linear elastic shell, while the fold zones were assigned material properties of spring steel to replicate hinge behavior. To enable self-deployment, Panel 3 was rotated quasi-statically relative to Panel 4, inducing folding and stress storage. The folded state was then released to observe autonomous deployment. Fig.~\ref{fig:abaqus_model} shows the pre-folded, folded, and deployed configurations of the ABAQUS model.

ABAQUS was employed with a dynamic explicit solver to handle nonlinear transient response. The mesh used refined quadrilateral elements concentrated at fold regions to resolve steep strain gradients. Convergence tests confirmed the stability of the chosen discretization. To ensure physical relevance, the FE model was validated against the experimental data of \cite{fulton2022deployment}, who performed gravity-offloaded tests on a four-panel Miura prototype with tape-spring hinges. Fold regions were provided with different thicknesses, as tape-spring hinges were present only between panels 3 and 4, and the rest of the hinges were fully compliant. 

\subsection{Equivalent Hinge Modelling}
Explicit modeling of tape-springs is computationally expensive. Instead, an energy-equivalent thin elastic hinge was derived based on \cite{seffen1999deployment}. The strain energy of a tape-spring during bending is expressed as
\begin{equation}
U_{ts} = \frac{D(1+\nu)}{R^2} R^2{\alpha} \theta,
\end{equation}
with $D=Et^3/[12(1-\nu^2)]$. Equating this with the bending energy of a uniform strip hinge,
\begin{equation}
U_{h} = \frac{D_w}{R^2} L w \theta, \quad D_w = \frac{E t_w^3}{12(1-\nu^2)},
\end{equation}
leads to
\begin{equation}
t_w = \left(\frac{t^3 (1+\nu) R^2{\alpha}}{L w}\right)^{1/3}.
\end{equation}
For the experimental setup (R = \SI{15.875}{mm}, L = \SI{292}{mm}, $R\alpha$ = \SI{35}{mm}, $\nu=0.35$, $t=0.2$ mm), using an effective hinge width $w=36.5$ mm, the equivalent hinge thickness was $t_w\approx0.108$ mm. This equivalence allows computational efficiency while retaining strain-energy fidelity. The rest of the hinges were provided with minimal thickness to maximize compliance.

\subsection{Simulation Results}
\begin{figure}[h]
    \centering
    \includegraphics[width=1.2\linewidth]{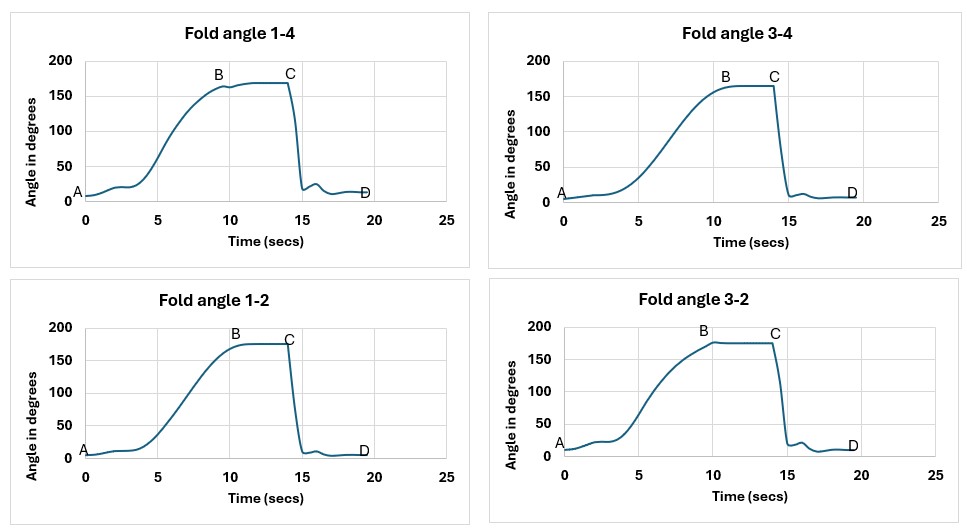}
    \caption{Variation of fold angles between panels with time during folding (A-B), during release of stresses (B-C), and deployment stage (C-D).}
    \label{fig:panel_angles}
\end{figure}

\begin{figure}[h]
    \centering
    \includegraphics[width=1.2\linewidth]{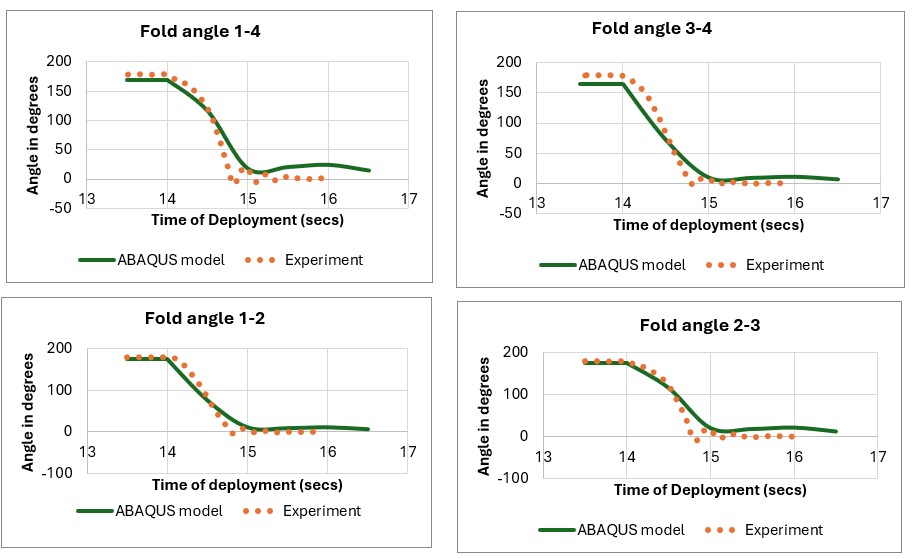}
    \caption{Comparison of deployment fold angles between simulation and experiment (\cite{fulton2022deployment}).}
    \label{fig:fold_comparison}
\end{figure}

The FE simulation reproduced deployment within 2.5 seconds, compared to 2.0 seconds observed experimentally. The variation of the fold angles through the process of folding, stress release, and deployment is shown in Fig.~\ref{fig:panel_angles}. The orientation of fold angles over time during the deployment phase (Fig.~\ref{fig:fold_comparison}) matched well with experimental data, with only minor oscillatory discrepancies. These differences were attributed to damping assumptions and hinge idealizations, particularly the continuous thin-shell approximation of tape-springs. Despite this, the FE model demonstrated strong predictive capability for both folding and deployment phases.

\subsection{Parametric Studies}
\begin{figure}[h]
    \centering
    \includegraphics[width=0.8\linewidth]{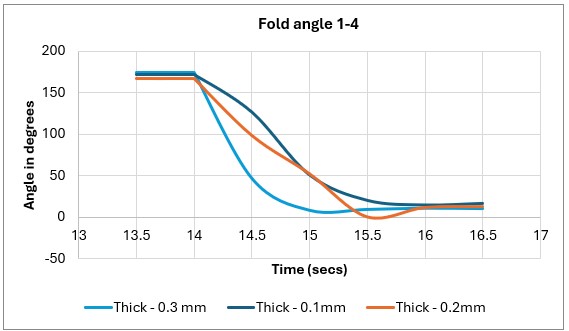}
    \caption{Effect of tape-spring equivalent thickness on deployment trajectory.}
    \label{fig:hinge_thickness}
\end{figure}

\begin{figure}[h]
    \centering
    \includegraphics[width=0.8\linewidth]{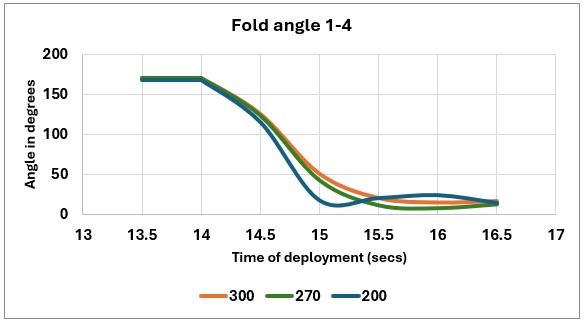}
    \caption{Influence of mass-damping coefficient $\alpha$ on deployment trajectory.}
    \label{fig:damping_effects}
\end{figure}

To assess sensitivity of deployment, the thickness of the fold region and the damping coefficients were varied:
\begin{itemize}
    \item \textbf{Hinge thickness:} As evident from Fig.~\ref{fig:hinge_thickness}, increasing equivalent hinge thickness accelerated deployment. For example, a thickness of 3 mm reduced deployment time to 1 s, demonstrating stiffness-dependent response.
    \item \textbf{Damping:} Rayleigh damping coefficient associated with the mass matrix $\alpha$ was varied to observe its effect. According to Fig~\ref{fig:damping_effects}, higher damping reduced oscillations and promoted faster attainment of steady deployed states, while low damping introduced prolonged oscillatory settling.
\end{itemize}

These studies reveal that deployment performance is highly tunable by adjusting hinge stiffness and damping properties, offering valuable design flexibility for aerospace applications.

\section{SMP-Based Deployment in PLA Miura-column}
\begin{figure}[h]
    \centering
    \includegraphics[width=1.1\linewidth]{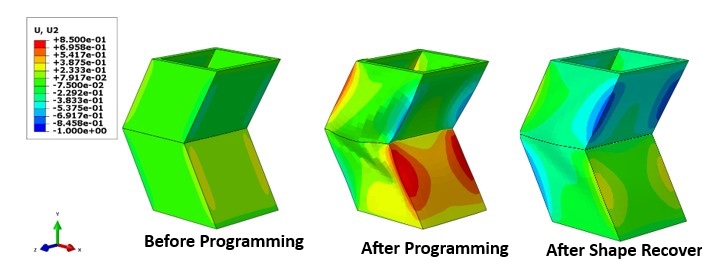}
    \caption{Displacement contours of PLA origami column during programming and recovery.}
    \label{fig:shapememorycolumn}
\end{figure}

Compressive tests on a PLA-based origami column (Fig.~\ref{fig:shapememorycolumn}), following the experimental studies by \cite{liu2018shape}, were chosen to simulate shape-memory-polymer-based deployment. The column structure was subjected to quasistatic compression at \SI{90}{\celsius}, cooled to retain the deformed shape, and reheated to recover its original configuration. The thermo-mechanical constitutive behaviour of the SMP was modelled using the two-temperature viscoelastic theory \cite{das2018constitutive}, implemented in a VUMAT subroutine.

\begin{figure}[h]
    \centering
    \includegraphics[width=1.2\linewidth]{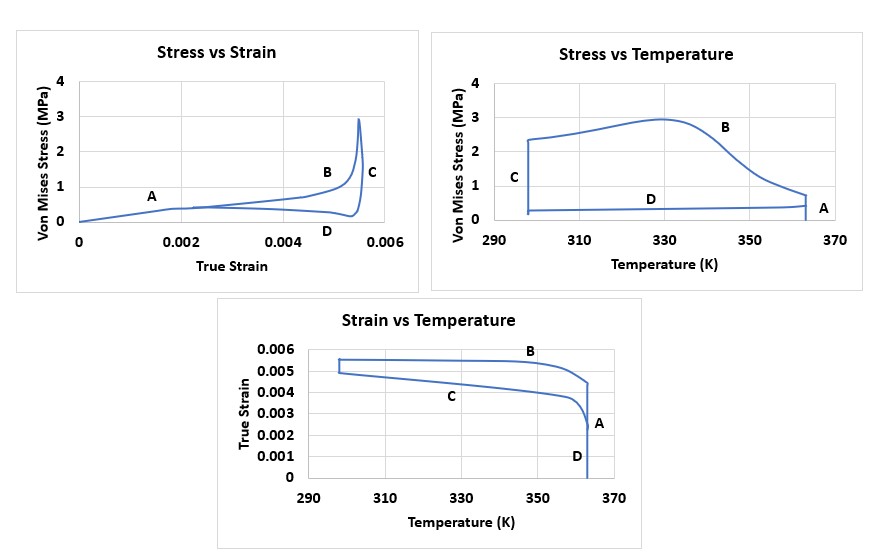}
    \caption{Stress-strain-temperature response of SMP-based Miura column through (A) Deformation at 90$^{\circ}$C, (B) Cooling down to 25$^{\circ}$C, (C) Release of stresses at 25$^{\circ}$C, and (D) Strain release at 90$^{\circ}$C.}
    \label{fig:pla_contours}
\end{figure}

Simulation results indicated a shape recovery ratio of 34.5\%. The stress-strain-temperature curves confirmed the expected thermomechanical transitions of PLA. Displacement contours (Fig.~\ref{fig:pla_contours}) illustrated programming, freezing, and recovery phases. Although partial recovery was achieved, the model captured key trends observed in experimental work, supporting the applicability of SMPs for active origami deployment.

\section{Conclusion}
This study developed finite element frameworks for simulating Miura-origami self-deployable structures through two complementary mechanisms: elastic strain energy release and shape-memory polymer (SMP) activation. For elastic hinge-driven deployment, the model was validated against experimental data, showing strong agreement in fold trajectories and deployment timing. The use of equivalent hinges enabled computational efficiency while preserving strain-energy fidelity, and parametric studies highlighted the sensitivity of deployment performance to hinge stiffness and damping.

SMP-based simulations demonstrated that a thermoviscoelastic constitutive model could capture the key features of programming, freezing, and shape recovery in PLA origami columns, producing realistic recovery ratios and thermomechanical trends. Together, these results confirm that mechanics-based finite element models are capable of providing accurate, predictive insights for the design of origami-inspired deployable structures in aerospace applications.

Future work should focus on experimental validation of SMP-driven deployment, optimization of hinge designs for robustness, and the integration of multiphysics effects, such as thermal gradients and orbital loading, to advance the design of autonomous self-deployable structures.


\begin{thebibliography}{0}
\expandafter\ifx\csname natexlab\endcsname\relax\def\natexlab#1{#1}\fi
\providecommand{\url}[1]{\texttt{#1}}
\providecommand{\href}[2]{#2}
\providecommand{\path}[1]{#1}
\providecommand{\DOIprefix}{doi:}
\providecommand{\ArXivprefix}{arXiv:}
\providecommand{\URLprefix}{URL: }
\providecommand{\Pubmedprefix}{pmid:}
\providecommand{\doi}[1]{\href{http://dx.doi.org/#1}{\path{#1}}}
\providecommand{\Pubmed}[1]{\href{pmid:#1}{\path{#1}}}
\providecommand{\bibinfo}[2]{#2}
\ifx\xfnm\relax \def\xfnm[#1]{\unskip,\space#1}\fi

\end{thebibliography}


\begin{thebibliography}{8}


\bibitem{fulton2022deployment} 
Fulton, R., Smith, J. and Lee, K., 2022. Experimental investigation of Miura-origami deployable structures. \textit{Journal of Mechanical Design}, 144(6), p.061702. https://doi.org/10.1115/1.4054115

\bibitem{seffen1999deployment} 
Seffen, K.A. and Pellegrino, S., 1999. Deployment dynamics of tape springs. \textit{Proceedings of the Royal Society A: Mathematical, Physical and Engineering Sciences}, 455(1983), pp.1003--1048. https://doi.org/10.1098/rspa.1999.0340

\bibitem{liu2018shape} 
Liu, Y., Wang, H. and Song, Z., 2018. Shape memory polymer origami structures: Experimental characterization and applications. \textit{Smart Materials and Structures}, 27(3), p.035008. https://doi.org/10.1088/1361-665X/aa9b22

\bibitem{das2018constitutive} 
Das, S. and Kumar, N., 2018. Thermoviscoelastic constitutive model for shape memory polymers. \textit{International Journal of Solids and Structures}, 147-148, pp.136--148. https://doi.org/10.1016/j.ijsolstr.2018.06.017

\bibitem{smith2020abaqus} 
Smith, M., 2020. \textit{ABAQUS/Standard User's Manual, Version 2020}. Dassault Systèmes Simulia Corp, United States.


\end{thebibliography}
\end{document}